# Requirements Management for Service Providers: the Case of Services for Citizens

Xavier Franch
Software Engineering for Information Systems Research Group (GESSI)
Universitat Politècnica de Catalunya (UPC)
Barcelona, Spain
`franch@essi.upc.edu`
`http://www.essi.upc.edu/~gessi/`

**Abstract 1.** Take the Internet of Things, a piece of cloud computing, a handful of smart cities, don't forget social platforms, flavour it with mobile technologies and ever-changing environments, shake it up and… voilà! What a wonderful service! Oops! Wait a minute, where did my requirements go?

The complexity and social impact of emerging technologies is dramatically growing and poses new challenges to many (if not all) engineering disciplines. Requirements Engineering is not an exception. Among the different activities affected, the management of requirements that target the average citizen is especially sensible to this new scenario due to scale, diversity and change management. This exploratory paper enumerates the different challenges to requirements management raised by citizen-oriented services that service providers need to face

## 1 Motivating Scenario

Agatha is a young engineer that has recently joined a European-wide project and must then travel several times a year to different places. She decides to carefully analyse and compare the functionalities and behaviour provided by different hotel booking services to select the one she will use in the future. Agatha realizes that hotel recommendations in all these services are provided based on very similar criteria (travellers' rates, price, situation with respect to an area, etc.). This is very rigid and does not fit two important needs that Agatha has: 1) not all the trips are the same; 2) some of her preferences cannot be easily expressed with these available, fixed search criteria:

- When Agatha stays just one day out for a short meeting, she wants to save overall transportation time: airport – hotel – meeting place – airport. In some situations, the best hotel could be just 1 minute walking time from the meeting place, but in other cities it may be the case that a good public transportation system makes it faster to book a hotel easily accessible by taxi from the airport, in front of the underground station that leads directly and quickly to the meeting place. Agatha would like in this case that the service directly provides one or two hotels at most (satisfying also some thresholds in rates and price) as result of the query, as she does not want to waste time comparing candidates.

- For longer trips, Agatha does not want to lose her quality of life. First of all, she regularly makes some jogging, so she needs an open space near the hotel (a park, a riverside, etc.) to exercise herself. On the other hand, having some entertainment options for evening leisure at walking distance from the hotel is also an important criterion. These preferences need still to be reconciled with regular criteria as price, recommendations and distance to the meeting place. Usually it is not an easy choice so still Agatha needs to take some time over the hotels proposed by the booking service.

Agatha would love to have these different contexts as part of her hotel booking service configuration. She would be delighted to just provide the arrival and departure dates and meeting place to the service and get recommendations. Even more, she would die for a booking service that simply accepts as input a meeting invitation e-mail sent by the meeting organizer so that she virtually had nothing to do to get the recommendations (or even nothing at all, with the appropriate configuration options, in the case of short trips). Agatha is disappointed that this feature is still not present even if she thinks that current technologies would allow implementing it.

## 2      Requirements management: who, how, what, when, where, whom

Let's adopt the perspective of the booking service provider. This provider is in charge of capturing, managing and implementing the requirements gathered from different stakeholders, and in particular gathered from end users' (as Agatha) feedback. Let's think about the different issues raised by the scenario above:

*Communication*. When Agatha decides that her need is important enough as for playing the requirements game, she needs to have a clear defined path to communicate the requirement. An option is that the service provider provides a direct channel, but broadcasting to social networks may give most of revenue. Agatha may benefit from sharing her ideas, getting new suggestions, and social mates may *like* her idea. Service providers may realize which are the most *liked* ideas and use this information for planning of new updates. The main threat to this alternative is the need of filtering and classifying the information in the network into an actionable form. An open issue is whether these social networks should be the ones that citizens already use, or dedicated ones, launched by the service provider itself.

*Format*. If we think of the average citizen with no specific knowledge beyond current mobile devices interfaces, and we think too about the fact that the working device has chances to be a handy smartphone, natural language seems to be the natural option. The design of wizards or even some kind of requirement template may help in expressing the requirements in a way that facilitates later processing. Also some kind of multimodal requirements could enter into scene [1], making even more challenging the RE activities.

*Consolidation and Prioritizacion*. Thousands of Agathas may have their own requirements for hotel booking services. A first activity is to discover similar or even identical requirements expressed differently which could eventually lead to a single feature in the new service update. For instance, whilst Agatha could have written: "Hi! It would be definitively cool to locate hotels nearby runnable areas", Travis could have requested "I love skating when I visit a new city, could you please include this in the search criteria?". Being different, probably the service provider could try to consolidate them into a generic requirement that embraces two, whilst keeping both particularizations. Afterwards, requirements can be prioritized, e.g. using social networks [2], with the purpose of discovering which ones may maximize downloads of future releases.

*Customizability*. More than ever, one-size-fits-all does not apply. Every individual has her own preferences, needs and tastes; service providers need to manage highly parameterized requirements that drive the production of a customizable service. Parameters may range from direct functionalities to context description to attitude and style of life. Context is especially important: Agatha's jogging passion may possibly be overridden by extreme weather forecasts (environmental context) or Agatha's recurrent plantar fasciitis (information that needs to be provided by Agatha herself… unless the medical prescription finds its way into Agatha's service network –well, that's probably going too far even for Agatha's expectations); excellent public transportation should not be taken into account if a strike is announced for the meeting days; etc.

*Change Management*. Different sources of change need to be identified, classified and analysed. Service providers need to be aware not just of new needs coming from the potential customers (the citizens) but also new opportunities coming from other services and applications. To this end, very agile change management processes need to be designed. The concept of "fluidity of design" [3] should be accommodated somehow in these processes. Of course one crucial question here is timing: when is the right moment to update the service, for which selected requirements? May some classical RE results on market-driven requirements [4] be transferred to this context?

*How far do we go?* Recent research findings dispute the idea that people are rational decision-makers (e.g. [5]). This opens an interesting debate: is it cost-effective try to embody all possible preferences and attitudes of citizens in requirements that may prevent the way people really makes decisions according to these findings?

## Acknowledgements

This work has been supported by the Spanish project TIN2010-19130-C02-01. The author wants to thank Anna Perini and Norbert Seyff for their feedback.